\documentclass{article}

\usepackage[T1]{fontenc}
\usepackage{lmodern}

\usepackage{amsmath,amssymb,amsfonts,amsthm}
\usepackage{mathtools}
\usepackage{graphicx}
\usepackage{tikz}
\usepackage{float}
\usepackage{bm}
\usepackage{url}
\usepackage{underscore}
\usepackage{colortbl}
\usepackage{tabularx}
\usepackage{fancyhdr}
\usepackage[normalem]{ulem}
\usepackage{textcomp}
\usepackage{xcolor}
\usepackage{ascmac}
\usepackage{multicol}
\usepackage{wrapfig}
\usepackage{algorithmic}
\usepackage{appendix}
\usepackage{comment}
\usepackage[utf8]{inputenc}
\usepackage[T1]{fontenc}
\usepackage{booktabs}
\usepackage[left=30mm,right=30mm]{geometry}
\usepackage{booktabs}
\usepackage{enumitem}

\usepackage{subcaption}

\theoremstyle{definition}

\newtheorem{theorem}{Theorem}
\newtheorem{lemma}{Lemma}

\newtheorem{definition}{Definition}

\newcommand{\btcl}{\textsc{B2CL}}
\newcommand{\pak}{\textsc{PArcK}}
\newcommand{\ak}{\textsc{Arc Kayles}}
\newcommand{\gh}{\textsc{Generalized Hex}}

\newcommand{\blue}{\textit{Blue}}
\newcommand{\red}{\textit{Red}}

\newcommand{\craig}[1]{\color{orange}#1\color{black}}

\title{Planar PArcK is PSPACE-complete}
\author{Craig Tennenhouse, Kanae Yoshiwatari}
\date{}

\begin{document}

\maketitle
\begin{abstract}
\textsc{Partizan Arc Kayles}, or \pak, is a partizan combinatorial game on edge colored graphs, and is a generalization of the game \textsc{Domineering}. Previously, \pak\ was shown to be PSPACE-complete on bipartite graphs. In this paper we show that \pak\ is PSPACE-complete even when restricted to planar graphs of degree at most three by a reduction from \gh.
\end{abstract}

\section{Introduction}

\textsc{Partizan Arc Kayles} (\pak) is played on an undirected graph in which every edge is colored either blue or red.
Players can choose an edge with their color on their turn, and its endpoints and all their incident edges are removed.
The player who has no move on their turn loses.

\pak\ is often studied due to its relevance as a generalization of \textsc{Domineering}, the latter being of special interest due to a recent result using OpenAI~\cite{mazur2026exact} to discover a counterexample to a longstanding conjecture of Berlekamp's regarding the temperature of the ruleset (referenced in~\cite{Berlekamp2019}). \pak\ on general graphs was shown to be NP-hard in~\cite{hanaka2025faster} by a reduction from \textsc{Vertex Cover}, then later shown in~\cite{burke2026itshardparck} to be PSPACE-complete even when restricted to bipartite graphs. 
\ak\ is an impartial version of \pak\ and was recently shown to be PSPACE-complete~\cite{bonnet2026arckaylespspacecomplete}.

In this paper, we show that \pak\ is PSPACE-complete even on planar graphs with degree at most three by means of a reduction from \gh. 
Typically planar PSPACE-completeness is addressed via a reduction from \textsc{Bounded 2-Player Constraint Logic} (\btcl)~\cite{hearn2009games}. 
We also show that a certain gadget required for a reduction from \btcl\ cannot be constructed for \pak\ under a natural method that represents the orientations of \btcl\ edges by maximum matchings within the gadget.

\section{Preliminaries}

\subsection{Related works}
\begin{figure}[ht]
\centering
\includegraphics[width=0.5\textwidth]{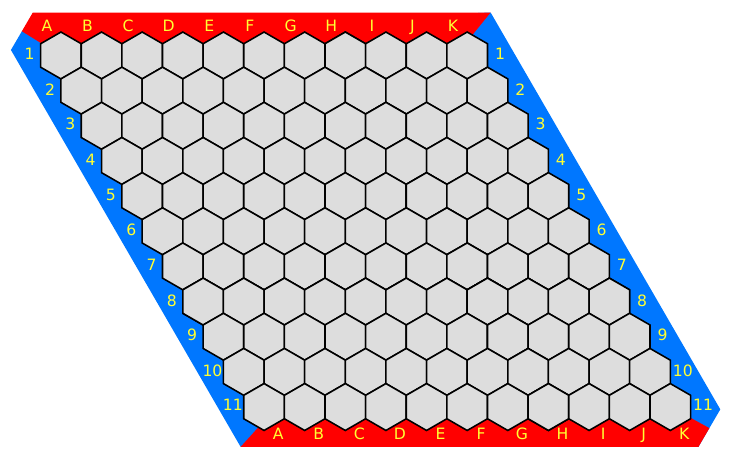}
\caption{A standard $11\times 11$ \textsc{Hex} board, Public domain, via Wikimedia Commons}
\label{fig:hexboard}
\end{figure}

\textsc{Hex} is a popular game commonly played on an $n\times n$ board with hexagonal spaces, where $n$ is usually equal to $11$ (see Figure \ref{fig:hexboard}). Players alternately place stones of their color and the first player to create an unbroken path of stones of their color between their two specified sides of the board is the winner. Thus \textsc{Hex} is usually played using \emph{affine} normal play rules, i.e. there is a win condition that can be met before the board is exhausted. Other affine games include \textsc{Chess} and \textsc{Connect Four}. \gh\ is a generalization of \textsc{Hex} to undirected graphs. In \gh\ a connected graph $G$ has two special vertices, $s$ and $t$. Players alternate coloring uncolored nodes from $G\setminus\{s,t\}$. Once all playable vertices are colored, \blue\ wins if there is a path of only blue vertices from $s$ to $t$, and \red\ wins otherwise. The authors of ~\cite{even1976combinatorial} proved that \gh\ is PSPACE-complete on general graphs, and in~\cite{reisch1981hex} Reisch showed that is PSPACE-complete even when the underlying graph is planar and the vertices $s$ and $t$ are in the same planar face. It is this latter result that we denote by \gh\ in this paper.

\subsection{Notations}
For a graph $G$, we denote by $V(G)$ and $E(G)$ the vertex set and the edge set of $G$, respectively.
For a vertex $v \in V(G)$, we denote by $d(v)$ the degree of $v$.
For a vertex set $S \subseteq V(G)$, we denote by $G \setminus S$ the graph obtained from $G$ by removing the vertices in $S$ and their incident edges.
A \emph{matching} of $G$ is a set $M \subseteq E(G)$ of edges no two of which share an end.
We denote by $V(M)$ the set of ends of the edges in $M$,
and say that $M$ \emph{covers} a vertex $v$ if $v \in V(M)$.
A matching $M$ is \emph{perfect} if $V(M) = V(G)$,
and a \emph{maximum matching} of $G$ is a matching with the maximum number of edges.
For two matchings $M_1$ and $M_2$, we denote by $M_1 \triangle M_2$ their symmetric difference.
For two graphs $G_1$ and $G_2$ with no common vertex, we denote by $G_1 \sqcup G_2$ their disjoint union.

\section{PSPACE-completeness of PArcK on planar graphs}

\begin{figure}[ht]
\centering
\includegraphics[width=0.5\linewidth]{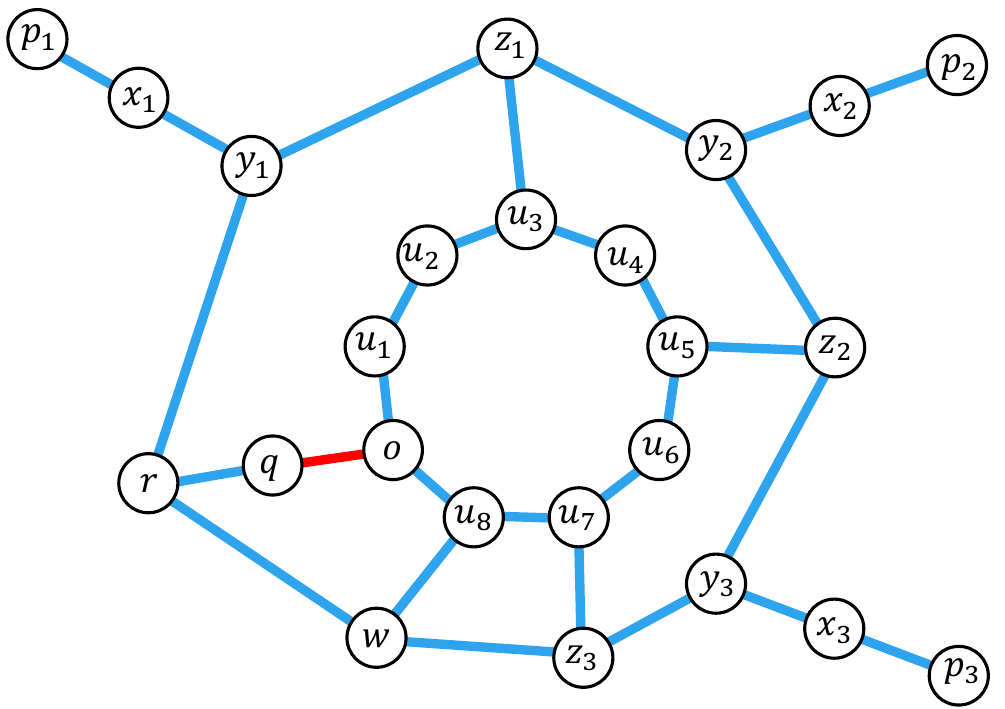}
\caption{A graph $C$ with $d=3$}
\label{fig:vg}
\end{figure}

We reduce from \gh\ on planar graphs.
We first introduce Lemmas~\ref{lem:blue}–\ref{lem:red}, which guarantee the number of moves that Blue can make in a graph $C$. In our reduction, we replace each vertex of the $\gh$ instance with a copy of $C$. For an integer $d \geq 2$, we construct $C$ as illustrated in Figure~\ref{fig:vg}. The vertex set $V(C)$ and the edge set $E(C)$ are defined as follows:

\begin{align*}
V(C)={}&\bigcup\limits_{i=1}^{d} \{p_i,x_i,y_i,z_i\} 
\cup \bigcup\limits_{i=1}^{2d+2} \{u_i\} 
\cup \{o,q,r,w\},
\end{align*}

\begin{align*}
E(C)={}&\bigcup\limits_{i=1}^{d}
\left\{ \{p_i,x_i\},\{x_i,y_i\},\{y_i,z_i\}\right\} \cup \bigcup\limits_{i=1}^{d-1} \bigl\{\{z_i,y_{i+1}\}\bigr\} 
\cup \bigcup\limits_{i=1}^{2d+1} \bigl\{\{u_i,u_{i+1}\}\bigr\} \cup \bigcup\limits_{i=1}^{d} \bigl\{\{z_i,u_{2i+1}\}\bigr\} \\
&\cup \left\{
\{o,q\},\{r,q\},\{r,w\},
\{o,u_1\},\{o,u_{2d+2}\},
\{w,z_d\},\{w,u_{2d+2}\},\{r,y_1\}
\right\}.
\end{align*}

We refer to the $p_i$ as \emph{ports}. In $E(C)$, only $\{o,q\}$ is red, and all other edges are blue.
The graph $C$ has $|V(C)| = 6d + 6$ vertices.

\begin{lemma}\label{lem:blue}
The graph $C \setminus \{q,r\}$ has a perfect matching. When we remove any two $p_i$ from $C \setminus \{q,r\}$, it still has a perfect matching.
\end{lemma}

\begin{figure}[ht]
\centering
\begin{subfigure}{0.48\linewidth}
    \centering
    \includegraphics[width=\linewidth]{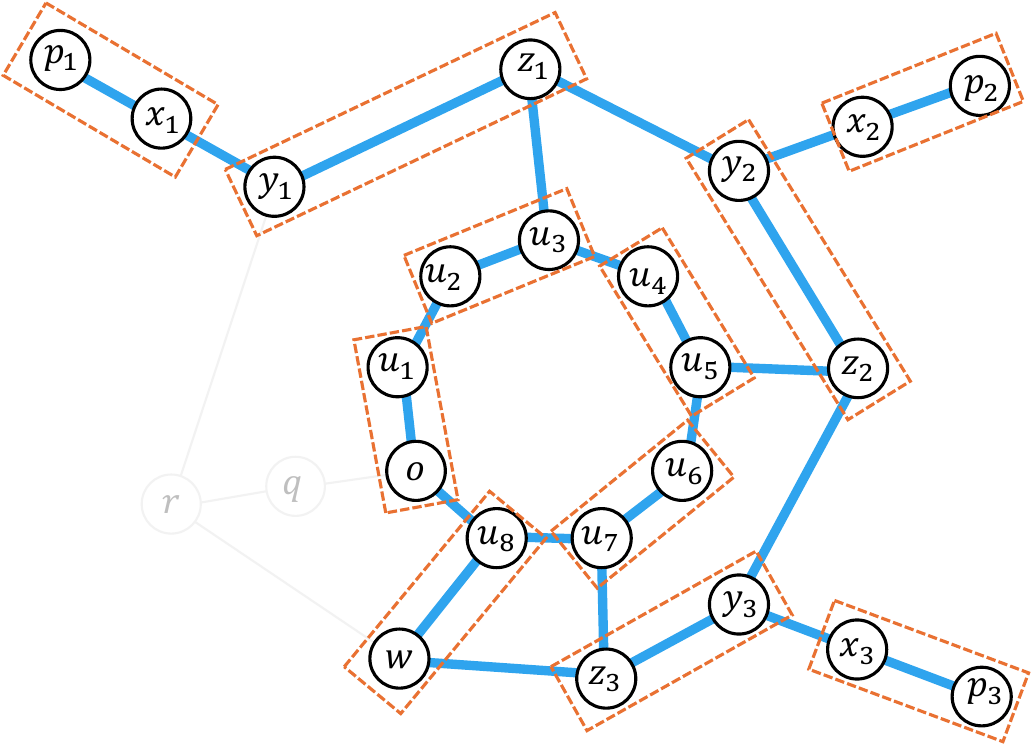}
    \caption{$C\setminus \{q,r\}$}
    \label{fig:vg_blue}
\end{subfigure}
\hfill
\begin{subfigure}{0.48\linewidth}
    \centering
    \includegraphics[width=\linewidth]{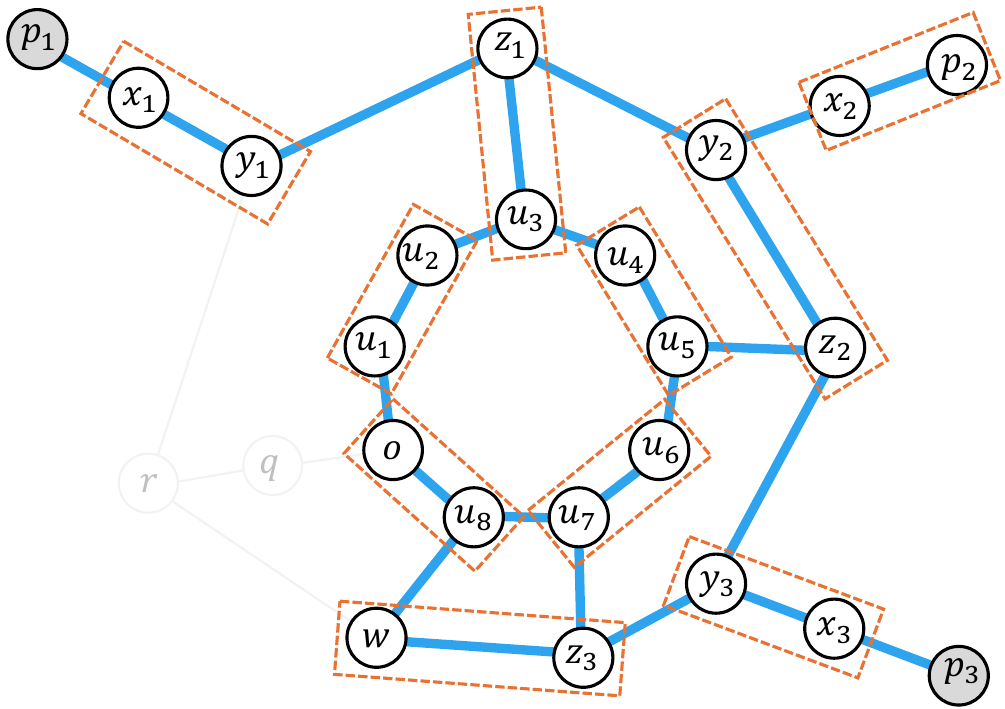}
    \caption{$C\setminus \{q,r,p_a,p_b\}$ ($a=1,b=3$)}
    \label{fig:vg_blue_port}
\end{subfigure}
\caption{Perfect matchings}
\label{fig:vg_blue_all}
\end{figure}

\begin{proof}
For $C \setminus \{q,r\}$, we consider a matching $M$ shown in Figure~\ref{fig:vg_blue}. $M$ consists of $\{p_k,x_k\}$ for $k\in \{1,\ldots, d\}$, $\{y_k,z_k\}$ for $k\in \{1,\ldots, d\}$, $\{w,u_{2d+2}\}$, $\{o,u_{1}\}$, and $\{u_{2k},u_{2k+1}\}$ for $k\in \{1,\ldots, d\}$.
Since $|V(C \setminus \{q,r\})|=6d+4$ and $|M|=3d+2$, $M$ is a perfect matching on $C \setminus \{q,r\}$.

For any $1 \leq a < b \leq d$, we next consider $C \setminus \{q,r,p_a,p_b\}$ and a matching $M$ on it. See Figure~\ref{fig:vg_blue_port}.
$M$ consists of $\{x_a,y_a\}$, $\{x_b,y_b\}$, $\{z_a,u_{2a+1}\}$, $\{p_k,x_k\}$ for $k \in \{1,\ldots,d\}\setminus\{a,b\}$, $\{y_k,z_k\}$ for $k \in \{1,\ldots,b-1\}\setminus\{a\}$, $\{z_k,y_{k+1}\}$ for $k \in \{b,\ldots,d-1\}$, $\{z_d,w\}$, $\{o,u_{2d+2}\}$, $\{u_{2k-1},u_{2k}\}$ for $k \in \{1,\ldots,a\}$, and $\{u_{2k},u_{2k+1}\}$ for $k \in \{a+1,\ldots,d\}$.
Since $|V(C \setminus \{q,r,p_a,p_b\})|=6d+2$ and $|M|=3d+1$, $M$ is a perfect matching on $C \setminus \{q,r,p_a,p_b\}$.
\end{proof}

By arguments similar to that in Lemma~\ref{lem:blue}, Lemmas~\ref{lem:blue2} and~\ref{lem:blue3} hold for the other two blue edges adjacent to the red edge.

\begin{lemma}\label{lem:blue2}
The graph $C \setminus \{o,u_1\}$ has a perfect matching.
When we remove at least one $p_i$ from $C \setminus \{o,u_1\}$, it has no perfect matching.
\end{lemma}

\begin{lemma}\label{lem:blue3}
The graph $C \setminus \{o,u_{2d+2}\}$ has no perfect matching.
When we remove any two $p_i$ from $C \setminus \{o,u_{2d+2}\}$, it has a perfect matching.
\end{lemma}

Lemma~\ref{lem:red} holds for the graph obtained after \red\ selects the red edge in $C$.

\begin{lemma}\label{lem:red}
The graph $C \setminus \{q,o\}$ has a perfect matching. When we remove at least one $p_i$ from $C \setminus \{q,o\}$, it has no perfect matching.
\end{lemma}

\begin{figure}[ht]
\centering
\includegraphics[width=0.5\linewidth]{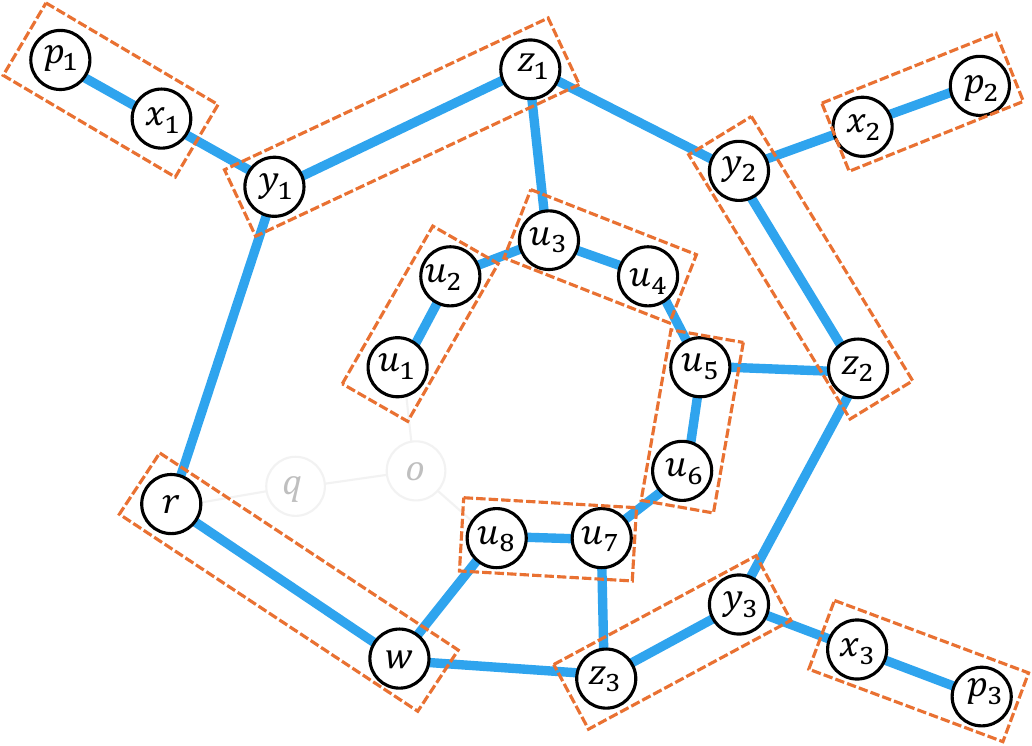}
\caption{A perfect matching of $C\setminus \{q,o\}$}
\label{fig:vg_red}
\end{figure}

\begin{proof}
For $C \setminus \{q,o\}$, we consider a matching $M$ shown in Figure~\ref{fig:vg_red}. $M$ consists of $\{p_k,x_k\}$ for $k\in \{1,\ldots, d\}$, $\{y_k,z_k\}$ for $k\in \{1,\ldots, d\}$, $\{w,r\}$, and $\{u_{2k-1},u_{2k}\}$ for $k\in \{1,\ldots, d+1\}$.
Since $|V(C \setminus \{q,o\})|=6d+4$ and $|M|=3d+2$, $M$ is a perfect matching on $C \setminus \{q,o\}$. 

$C \setminus \{q,o\}$ is bipartite and we see the partition $(V_1,V_2)$ as follows: $V_1$ includes $p_k$ for $k\in \{1,\ldots, d\}$, $y_k$ for $k\in \{1,\ldots, d\}$, $w$, and $u_{2k-1}$ for $k\in \{1,\ldots, d+1\}$, while $V_2$ includes $x_k$ for $k\in \{1,\ldots, d\}$, $z_k$ for $k\in \{1,\ldots, d\}$, $r$, $u_{2k}$ for $k\in \{1,\ldots, d+1\}$. 
When we remove at least one $p_k$, $|V_1|<|V_2|$ holds and there exists no perfect matching.
\end{proof}

\begin{lemma}\label{lem:st}
Any position of \gh\ can be transformed into a position with the same winner in which both $s$ and $t$ have degree at most two. This transformation preserves planarity.
\end{lemma}

\begin{proof}
Let $G$ be a position of \gh.
If $d(s) \le 2$, we do nothing for $s$.
Otherwise, let $v_1, v_2, \ldots, v_{d(s)}$ be the neighbors of $s$ in clockwise order around $s$.
We remove all edges incident to $s$, add a path $P = (s_1, s_2, \ldots, s_{d(s)})$ with $s_1 = s$,
and add the edge $\{s_i, v_i\}$ for each $i$.
All vertices of $P$ other than $s$ are colored blue at the beginning of the game.
We call the resulting graph $G'$.
Then $s$ has degree two in $G'$.
Since $P$ can be drawn inside a small disk around $s$ and the edges $\{s_i,v_i\}$ keep the clockwise order,
$G'$ is planar.

The positions $G$ and $G'$ are the same except for $P$, so every play in $G$ is also a play in $G'$ and vice versa.
Consider the same final coloring of $G$ and $G'$.
If $G$ has an $s$-$t$ path $(s, v_j, \ldots, t)$ whose internal vertices are blue,
then $(s_1, \ldots, s_j, v_j, \ldots, t)$ is such a path in $G'$, since $s_2, \ldots, s_j$ are blue.
Conversely, suppose that $G'$ has such a path $Q$, and let $s_j$ be the last vertex of $Q$ in $P$.
Then replacing the subpath of $Q$ from $s_1$ to $s_j$ with $s$ yields such a path in $G$.
Applying the same transformation to $t$, we obtain the desired position.
\end{proof}

Let $G$ be a position of \gh\ on a planar graph with two specified vertices $s$ and $t$,
and fix a plane embedding of $G$.
We assume that $s$ and $t$ are not adjacent.
We also assume that every vertex of $G$ other than $s$ and $t$ has degree at least two.
Indeed, a vertex $v \notin \{s, t\}$ of degree at most one is not an internal vertex of any $s$-$t$ path,
so putting a piece on $v$ does not affect any $s$-$t$ path.
Such a move is equivalent to passing, but passing is never advantageous in \gh.
Hence we may remove such vertices repeatedly without changing the winner.
Each vertex of $G$ other than $s$ and $t$ is either colored blue, colored red, or uncolored.

In our reduction to \pak, from $G$ we construct a graph $H = H' \sqcup K$ whose edges are colored blue or red,
where $K$ consists of isolated red edges and $H'$ is described below.

The graph $H'$ is constructed as follows.
For each vertex $v \in V(G) \setminus \{s, t\}$, we take a copy $C_v$ of the gadget $C$ with parameter $d = d(v)$,
and remove $q$ and $r$ from $C_v$ if $v$ is blue, and $o$ and $q$ if $v$ is red.
We associate the ports of $C_v$ with the edges incident to $v$
so that their cyclic order on the outer face of $C_v$ matches the clockwise order of the edges around $v$ in $G$.
The graph $H'$ consists of the vertices $s$ and $t$ and the gadgets $C_v$ for all $v \in V(G) \setminus \{s, t\}$,
after the removal above, together with the following blue edges.
For each edge $\{v, w\}$ of $G$ with $v, w \notin \{s, t\}$,
we add a blue edge joining the port of $C_v$ and the port of $C_w$ that correspond to $\{v, w\}$.
Similarly, for each edge $\{s, w\}$ of $G$,
we add a blue edge joining $s$ and the port of $C_w$ that corresponds to $\{s, w\}$,
and we add blue edges incident to $t$ in the same way.
Since the ports of each $C_v$ appear in the same cyclic order as the edges around $v$ in $G$,
the graph $H'$ is planar.

We next construct $K$.
Let
\[
  \kappa = \sum_{v \in V(G) \setminus \{s, t\}} (3 d(v) + 2).
\]
Then $K$ consists of $\kappa + 1$ isolated red edges if the number of uncolored vertices of $G$ is odd, and of $\kappa$ isolated red edges otherwise.
This completes the construction of $H$.
Clearly, the size of $H$ is linear in the size of $G$, and $H$ is planar.

\begin{figure}[ht]
\centering
\includegraphics[width=0.9\linewidth]{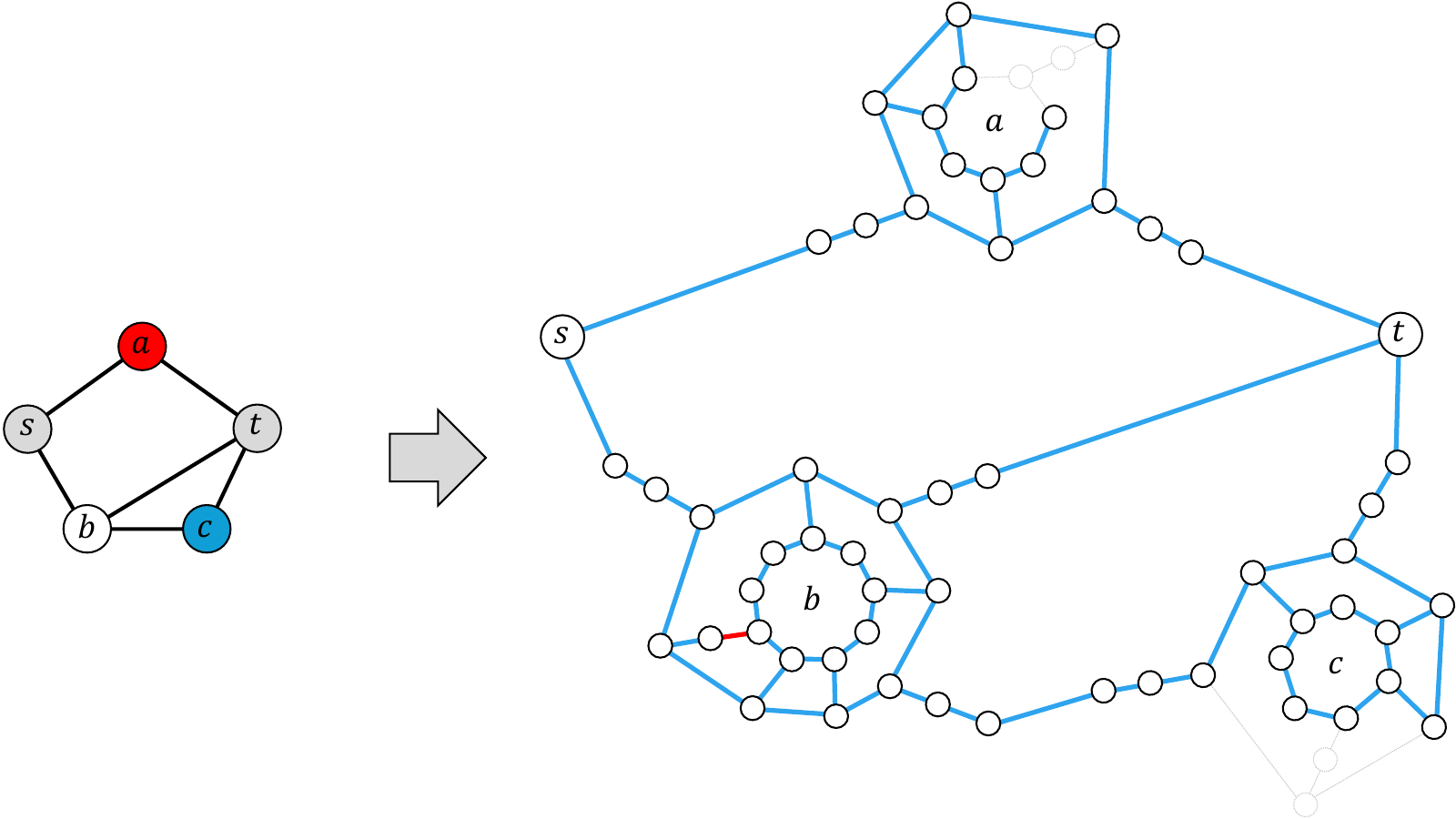}
\caption{An example of $H'$}
\label{fig:example}
\end{figure}

Note that under optimal play, while $H'$ has a red edge,
\blue\ chooses only blue edges that share a vertex with a red edge of $H'$,
and \red\ chooses only red edges of $H'$.
Intuitively, a blue edge that shares no end with a red edge cannot be removed by \red,
and an isolated red edge of $K$ cannot be removed by \blue.
Hence such an edge remains available until the end of the game, and choosing it while $H'$ has a red edge gives no advantage to the player who chooses it.

\begin{lemma}\label{lem:winner}
\blue\ wins \gh\ on $G$ if and only if \blue\ wins \pak\ on $H$.
\end{lemma}

\begin{proof}
Suppose that \blue\ has a winning strategy for \gh{} on $G$.
\blue\ simulates this strategy on $H'$:
whenever the strategy on $G$ chooses a vertex $v$, \blue\ chooses the edge $\{r,q\}$ in $C_v$ on $H'$.
Consider the moment when all uncolored vertices of $G$ have been colored.
At this moment, no red edge remains in $H'$ 
and $G$ contains an $s$-$t$ path $P = (s, v_1, v_2, \ldots, v_\ell, t)$ whose internal vertices are all blue.
From now on, \red\ can only move on isolated edges on $K$.

We now count the number of moves \blue\ can make on $H'$.
Given a matching $M$ of $H'$, \blue\ can make $|M|$ moves by choosing the edges of $M$ one by one.
We construct $M$ as follows.
\begin{enumerate}
  \item For each $i \in \{1, \ldots, \ell-1\}$, add to $M$ the edge joining the ports of $C_{v_i}$ and $C_{v_{i+1}}$.
  \item Add to $M$ the edge joining $s$ and a port of $C_{v_1}$, and the edge joining $t$ and a port of $C_{v_\ell}$.
  \item By Lemma~\ref{lem:blue}, the graph obtained from the current $H'$ by removing $V(M)$ has a perfect matching. Add it to $M$.
\end{enumerate}
The resulting $M$ is a perfect matching of the current $H'$.
Since each $C_v$ has $6d(v) + 6$ vertices before the removal and exactly two of them have been removed,
\[
  |M| = \frac{1}{2}\Bigl(2 + \sum_{v \in V(G) \setminus \{s, t\}} \bigl(6d(v) + 4\bigr)\Bigr)
      = 1 + \sum_{v \in V(G) \setminus \{s, t\}} \bigl(3d(v) + 2\bigr)
      = \kappa + 1.
\]

If the number of uncolored vertices of $G$ at the start of the game is odd,
then \red\ moves next at this moment and has $\kappa + 1$ isolated edges, so \blue\ wins.
If it is even, then \blue\ moves next and \red\ has $\kappa$ isolated edges, so \blue\ wins \pak{} on $H$.

Suppose that \red\ has a winning strategy for \gh\ on $G$.
While $G$ has an uncolored vertex, \red\ chooses the edge $\{o,q\}$ in $C_v$ for the vertex $v$ chosen by the winning strategy on $G$.
\blue\ chooses only the edges $\{q,r\}$, $\{o,u_1\}$, and $\{o,u_{2d+2}\}$ of some $C_v$ on $H'$.

Consider the moment when no red edge remains in $H'$.
By \red's strategy, $G$ then has no $s$-$t$ path consisting of blue vertices,
and the current $H'$ has $2\kappa + 2$ vertices as before.
Afterwards, the number of \blue's moves equals the size of a maximum matching of the current $H'$, which is at most $\kappa + 1$.

Suppose, for contradiction, that the current $H'$ has a perfect matching $M$.
Since every neighbor of $s$ is a port of some gadget, $s$ is matched by $M$ to a port of some $C_{v_1}$.
Consider the moment when we enter a port of $C_v$ along an edge of $M$.
Since no red edge remains in $H'$, the gadget $C_v$ is in one of the following three states.
\begin{itemize}
  \item $C_v \setminus \{q, r\}$ or $C_v \setminus \{o, u_{2d+2}\}$:
    since these graphs have an even number of vertices,
    removing the port we entered leaves an odd number of vertices.
    Hence another port of $C_v$ must be matched by $M$ to a vertex outside $C_v$,
    and we follow that edge to the next gadget.
  \item $C_v \setminus \{o, u_1\}$:
    the port we entered is matched by $M$ to a vertex outside $C_v$.
    However, by Lemma~\ref{lem:blue2}, $C_v \setminus \{o, u_1\}$ with at least one port removed has no perfect matching,
    a contradiction.
  \item $C_v \setminus \{q, o\}$:
    by Lemma~\ref{lem:red}, $C_v \setminus \{q, o\}$ with at least one port removed has no perfect matching, a contradiction.
\end{itemize}
Repeating this procedure, we either stop at a gadget in one of the last two states or eventually reach $t$.
In the latter case, all vertices traversed are blue,
which contradicts that $G$ has no $s$-$t$ path consisting of blue vertices.
Therefore the current $H'$ has no perfect matching, and \blue\ can make at most $\kappa$ moves from this moment on.
If the number of uncolored vertices of $G$ at the start of the game is odd,
then \red\ moves next at this moment and has $\kappa + 1$ isolated edges, so \red\ wins.
If it is even, then \blue\ moves next and \red\ has $\kappa$ isolated edges, so \red\ wins \pak\ on $H$.
\end{proof}

\begin{theorem}
    \pak\ is PSPACE-complete even when the graph is planar with maximum degree three.
\end{theorem}

\begin{proof}
\pak\ is in PSPACE, since each move removes at least one edge, so the length of a game is at most the number of edges, and the game tree can be searched in polynomial space.

For PSPACE-hardness, by Lemma~\ref{lem:winner}, \blue\ wins \gh\ on $G$ if and only if \blue\ wins \pak\ on $H$.
The graph $H$ is constructed from a position $G$ of planar \gh\ in polynomial time and is planar, as explained before.
Every vertex of $C$ has degree at most three, and every port has degree one in $C$.
Since each port is joined to exactly one other port or to one of $s$ and $t$, its degree in $H$ is two.
By Lemma~\ref{lem:st}, the degrees of $s$ and $t$ in $H$ are at most two.
Therefore $H$ has maximum degree three.
\end{proof}

\section{A reduction from \btcl}

PSPACE-completeness of \textsc{Col} and \textsc{Snort} on planar graphs is shown by a reduction from \btcl~\cite{burke2019pspace}.
In \btcl, we are given a directed graph in which each edge has weight $1$ or $2$ and is owned by \blue\ or \red.
Each vertex has an \emph{inflow constraint}, which is a lower bound on the total weight of its incoming edges.
We call an orientation of the edges \emph{legal} if it satisfies all inflow constraints, and \emph{illegal} otherwise.
The players alternately reverse one of their own edges at most once each so that the orientation remains legal.
The first player to reverse their own target edge wins.

In the reductions to \textsc{Col} and \textsc{Snort},
each edge of the constraint graph is represented by a pair of vertices,
and reversing the edge corresponds to choosing which vertex of the pair is played.
The reductions use five main gadgets, and each gadget has its own threshold:
if the orientations of the edges incident to the gadget are legal in \btcl,
\blue\ can make exactly as many moves in the gadget as its threshold,
whereas if they are illegal, \blue\ can make fewer moves than the threshold.
Summing up over all gadgets,
\blue\ can make $k$ moves if all the edge orientations are legal and \blue's target edge is reversed,
and \blue\ can make at most $k-1$ moves otherwise.

One might try to reduce \btcl\ to \pak\ in the same manner:
each edge of the constraint graph is represented by two edges of the \pak\ position,
its orientation corresponds to which of the two edges is chosen,
and the number of moves \blue\ can make in each gadget is estimated.
However, we show that an AND gadget cannot be constructed under this method.

\begin{definition}[AND structure]
An \emph{AND structure} of a constraint graph is the subgraph consisting of a vertex $u$ and its three incident edges,
namely two edges $e_1$ and $e_2$ of weight $1$ and one edge $f$ of weight $2$,
where the inflow constraint of $u$ is $2$.
Hence the inflow constraint of $u$ is satisfied if and only if
$f$ is directed into $u$, or both $e_1$ and $e_2$ are directed into $u$.
We say that an orientation of $e_1$, $e_2$, and $f$ is \emph{legal at $u$} if it satisfies this constraint,
and \emph{illegal at $u$} otherwise.
\end{definition}

\begin{definition}[AND gadget]
An \emph{AND candidate} in \pak\ is a graph $X$ consisting only of blue edges,
together with three distinct vertices $p_1$, $p_2$, and $q$,
which correspond to $e_1$, $e_2$, and $f$ of an AND structure, respectively.
For an orientation $\sigma$ of $e_1$, $e_2$, and $f$, let $X_\sigma$ be the graph obtained from $X$ by removing
$p_i$ if $e_i$ is directed away from $u$ for $i \in \{1, 2\}$, and $q$ if $f$ is directed away from $u$.
We say that $X$ has \emph{threshold} $c$ if
a maximum matching of $X_\sigma$ has exactly $c$ edges for every orientation $\sigma$ legal at $u$,
and at most $c - 1$ edges for every orientation $\sigma$ illegal at $u$.
An AND candidate with a threshold is called an \emph{AND gadget}.
\end{definition}

\begin{theorem}\label{thm:noAND}
There is no AND gadget in \pak.
\end{theorem}

\begin{proof}
Suppose, for contradiction, that there is an AND gadget $X$ with threshold $c$.
By the definition of an AND structure, a maximum matching has exactly $c$ edges in each of
$X$, $X \setminus \{p_1, p_2\}$, and $X \setminus \{q\}$,
and at most $c - 1$ edges in each of
$X \setminus \{p_1, q\}$, $X \setminus \{p_2, q\}$, and $X \setminus \{p_1, p_2, q\}$.

Let $M_A$ be a maximum matching of $X \setminus \{p_1, p_2\}$ and $M_B$ a maximum matching of $X \setminus \{q\}$,
so that $|M_A| = |M_B| = c$.
If $M_A$ does not cover $q$, then $M_A$ is a matching of $X \setminus \{p_1, p_2, q\}$ with $c$ edges, a contradiction.
Hence $M_A$ covers $q$.

Each component of the symmetric difference $M_A \triangle M_B$ is a path or an even cycle
whose edges alternate between $M_A$ and $M_B$.
Since $q$ is covered by $M_A$ but not by $M_B$, $q$ is an end of a path $P$ in $M_A \triangle M_B$.
Let $w$ be the other end of $P$.
If the edge of $P$ at $w$ belongs to $M_A$, then $P$ has one more edge of $M_A$ than of $M_B$,
and $M_B \triangle P$ is a matching of $X$ with $c + 1$ edges, a contradiction.
Otherwise, $P$ has the same number of edges of $M_A$ and $M_B$,
and $M' = M_A \triangle P$ is a matching with $c$ edges that does not cover $q$.
The vertices covered by $M'$ are those covered by $M_A$, except $q$, together with $w$.
Since $M_A$ covers neither $p_1$ nor $p_2$, the matching $M'$ is a matching of
$X \setminus \{p_1, p_2, q\}$ if $w \notin \{p_1, p_2\}$,
$X \setminus \{p_2, q\}$ if $w = p_1$, and
$X \setminus \{p_1, q\}$ if $w = p_2$.
In each case, $M'$ has $c$ edges, a contradiction.
\end{proof}

\section{Acknowledgments}

The authors worked with Claude Opus 5.5 from Anthropic to consider possible PSPACE-complete games from which to reduce. Once \gh\ was suggested, the authors prompted for a vertex gadget. The LLM provided an initial planar gadget with justification. The LLM-provided proof was verified. Then the gadget was modified by the authors to achieve the max degree three result and proven to satisfy the reduction. The proof of Theorem~\ref{thm:noAND} was also provided by Claude at the prompting of the authors.

The authors also wish to thank MEXT Supporting Pioneering Research through AI for 1,000 Discovery challenges Program (SPReAD) Japan Grant Number JPMXP1726301948 for their generous support.


\bibliographystyle{plain}
\bibliography{sankou}

\end{document}